Original Article

# Stress Drops Associated with Surface Crack Formation in Photo-aged Polypropylene during Three-Point Bending

[1]Kazuya Haremaki, [2]Yusuke Koide, [1]Takashi Uneyama,
[1]Yuichi Masubuchi, *[1,3]Takato Ishida

[1] *Department of Materials Physics, Nagoya University, Furo-cho, Chikusa, Nagoya 464-8603, Japan*

[2] *Graduate School of Engineering Science, The University of Osaka, 1-3 Machikaneyama, Toyonaka, Osaka 560-8531, Japan*

[3] *Institute for Advanced Research, Nagoya University, Furo-cho, Chikusa, Nagoya 464-8601, Japan*

*To whom correspondence should be addressed
ishida@mp.pse.nagoya-u.ac.jp

**ABSTRACT**

Using three-point bending, this study investigates surface-crack formation in photo-aged polypropylene (PP) that has a depth-dependent aging gradient. PP undergoes embrittlement under ultraviolet (UV) irradiation, and because the photo-oxidation proceeds inward from the irradiated surface, the embrittlement develops non-uniformly across the specimen thickness. PP specimens were mildly photo-aged by UV irradiation and had not yet developed visible surface cracks. Each specimen was bent in two configurations: with the UV-irradiated ("aged") surface on the tensile side, and with the opposite ("reverse") surface on the tensile side. When the aged surface was on the tensile side, the stress–strain curves exhibited several discrete stress drops, and *in-situ* side-view observation confirmed that the formation of each new surface crack coincided with a stress drop. In contrast, no clear stress drops were observed when the reverse surface was on the tensile side. These results show that the through-thickness gradient of embrittlement is directly reflected in the bending stress–strain response. Uniaxial tensile testing, the standard method for evaluating mechanical properties, formally assumes a nominally uniform deformation across the cross-section and therefore reflects the spatially averaged response. Three-point bending, by contrast, imposes the largest tensile strain at the specimen surface and thus selectively probes the embrittled surface layer, making it an effective method for detecting the surface embrittlement of photo-aged polymers.

KEYWORDS: Embrittlement, Photo-oxidation, Thickness, Diffusion-limited oxidation (DLO)

## 1. INTRODUCTION

During long-term use, commodity semicrystalline polymers such as polypropylene (PP) materials undergo changes in their structures and mechanical properties, most notably embrittlement[1–4]. Since these polymers are widely used in daily life, for example as construction materials and automotive components, evaluating their embrittlement is important. Plastics have also attracted attention as environmental pollutants. In particular, microplastics, generally defined as plastic fragments smaller than 5 mm[5], have become a serious environmental concern[6]. Previous studies have suggested that the embrittlement of plastics affects the generation of microplastics[7]. Therefore, understanding the embrittlement of degraded plastics is important not only from an engineering viewpoint but also from an environmental perspective.

Uniaxial tensile testing is one of the most common methods for evaluating the mechanical properties of polymer materials. It assumes a nominally uniform deformation over the cross-section and characterizes properties such as the Young's modulus, the yield stress, and the mechanical strength (ultimate stress). An unaged PP exhibits the ductile deformation behavior in which yielding, neck propagation, and strain hardening are clearly observed. The mechanical properties of photo-aged PP have also been evaluated by such tests[8–13)]. As photo-aging progresses, the fracture behavior changes from ductile to brittle[1,8,9,14)]. Such brittle fracture behavior originates from chemical structural changes caused by photo-oxidation.

Photo-oxidation is one of the major degradation processes induced by ultraviolet (UV) exposure. The first radicals are considered to arise from unstable species already present in the polymer—chiefly peroxides formed by a limited thermal oxidation of the surface during melt processing[15)] and residual catalyst species left from polymerization[16)]—which readily decompose to initiate the oxidation. In PP, the tertiary C–H bonds are preferentially attacked[17)] so that hydrogen abstraction leaves a macroradical (R·) on the chain. Once these macroradicals are generated, oxidation propagates through the mechanism known as autoxidation[18)]. During autoxidation, hydroperoxides and carbonyl groups are generated, both of which subsequently undergo photolysis under continued UV exposure. Photolysis of the hydroperoxides produces radical species that cause chain scission of the polymer, whereas photolysis of the carbonyls proceeds through Norrish type I and II reactions[19)]. Oxidation occurs particularly in the amorphous regions of the polymer. In semicrystalline polymers such as PP, this chain scission reduces the mechanical strength and simultaneously promotes chemi-crystallization[13,20)]. The combined effects of chain scission and chemi-crystallization lead to embrittlement and brittle fracture behavior.

Photo-aging generally progresses non-uniformly from the exposed surface, resulting in a depth-dependent distribution of aging states inside the polymer. Such heterogeneous aging structures strongly influence crack initiation and fracture behavior[11,13,21]. This non-uniform aging originates from both the attenuation of irradiation intensity and the diffusion and consumption of oxygen ($O_2$) inside the material[19,22]. The irradiation intensity decreases exponentially with depth from the exposed surface, following the Lambert–Beer law. In addition, the internal oxygen concentration is governed by diffusion from the surface, described by Fick's second law[23]. As a result of these effects, the surface region becomes more severely aged than the inner region, leading to heterogeneous mechanical properties across the specimen thickness[10]. The resulting aged surface layer has been reported to range in thickness from a few μm to 500 μm[24–26]. However, Schoolenberg and Vink reported that, although UV absorption also contributes to the shape of the degradation profile, the thickness of the degraded surface layer under UV irradiation is governed mainly by the limitation of oxygen diffusion[26].

Because photo-aging proceeds from the exposed surface, the surface layer is systematically the most embrittled region of the specimen, so that the initiation of surface cracks is governed by the embrittled state of this surface. Evaluating the embrittlement of the surface layer is therefore essential for understanding the cracking of photo-aged polymers. Once a crack forms, stress concentration at its tip can drive propagation and eventual fracture. From this perspective, conventional uniaxial tensile testing is not ideally suited to photo-aged PP. It samples the surface and the bulk equally, so it does not preferentially reflect the response of the thin embrittled surface layer that is of primary interest. To probe the effect of the aging gradient locally, methods such as nanoindentation have been employed[10,27]. In this study, we propose bending tests as an alternative method for evaluating the surface embrittlement of photo-aged polymers with heterogeneous aging structures.

Bending tests are also widely used for evaluating the mechanical properties of materials. In an ideal bending deformation, the specimen deforms along a curved profile, and the stress state continuously changes from compressive to tensile through the specimen thickness. The boundary between the compressive and tensile regions is referred to as the neutral surface. A particular advantage of bending tests for photo-aged materials is that the tensile strain (and hence the tensile stress) reaches its largest value at the specimen surface. Therefore, bending tests are expected to provide mechanical data that selectively reflect the surface properties of photo-aged materials.

The bending test is often used for the evaluation of the mechanical and fracture properties of heterogeneous structures such as coatings and sandwich structures[28,29]. In the sandwich structure, a soft core is bonded between stiff faces to raise the bending stiffness and strength at low weight[28]. Bending tests can be used to evaluate the delamination or crack resistance[30,31]. In particular, coated materials generally consist of layered structures with different mechanical properties between the coating layer and the substrate. Coating properties can be evaluated from the number and spacing of cracks[29], and analytical models based on simple approximations have been proposed to relate these crack patterns to the coating properties[32].

In this study, we apply three-point bending tests to photo-aged PP specimens in order to evaluate the mechanical response that reflects their surface embrittlement. Although three-point bending tests have previously been applied to photo-degraded PP[21,33], the influence of the depth-dependent aging distribution on the bending response has not yet been investigated. PP sheets were photo-aged from one side under a xenon lamp, so that a depth-dependent gradient of the aging state developed across the specimen thickness. To characterize the difference in aging state between the two surfaces, the aged surface and the reverse surface were separately analyzed by Fourier-transform infrared (FT-IR) spectroscopy. We focused on mildly photo-aged specimens that had not yet developed visible surface cracks before mechanical loading, so that crack formation could be monitored during the bending process itself. Each specimen was subjected to three-point bending under two opposite configurations: one in which the aged surface (the UV-irradiated face) was placed on the tensile side, and one in which the reverse surface (the less-aged opposite face) was placed on the tensile side. Because the largest tensile stress (and strain) in bending occurs at the outermost surface, these configurations allow the mechanical contribution of each surface to be probed selectively. The stress–strain curves obtained under the two configurations were compared, and the side surface of the specimen was monitored with a digital video camera during loading to relate the mechanical response to crack formation. In addition, the tensile-side surface of each specimen was observed by optical microscopy after the bending tests to confirm the formation and spatial distribution of cracks.

## 2. EXPERIMENTAL

### 2.1 Materials and Specimen Preparation

Isotactic PP pellets with a weight-average molecular weight of $M_w$ = 3.4 × $10^5$ and a polydispersity index of $M_w/M_n$ = 3.6 were purchased from Sigma-Aldrich. To prepare sheet specimens, the pellets were sandwiched between two polyimide-film-coated aluminum sheets (100 µm thick, 150 mm × 150 mm) together with a 1 mm-thick aluminum spacer (150 mm × 150 mm) having a 100 mm × 100 mm window cut at its center. The assembly was hot-pressed at 200 °C under 15 MPa for 15 min using a test press (MP-2FH, Toyo-Seiki, Japan). Immediately after molding, the sheets were quenched in ice water at 0 °C and were cut into rectangular bending specimens with dimensions of 30 mm × 3 mm × 0.8 mm (length × width × thickness) and dumbbell-shaped specimens conforming to JIS K6251 type 8 (equivalent to ISO 37 type 3) with gauge width and length of 4 mm × 10 mm. The specimens were subsequently annealed at 140 °C for 3 h in an ADP-200 vacuum oven (Yamato Scientific Co., Ltd., Japan). The crystallinity of the as-prepared specimens was determined to be 42 wt % by differential scanning calorimetry (DSC). This measurement was conducted by Discovery DSC 25 (TA Instruments, USA) at a heating rate of 10 °C/min. A reference enthalpy of fusion of 209 J/g for fully crystalline PP was used[34)]. The crystallinity (42 wt %) may appear somewhat low;

however, it depends on the measurement and analysis method. For example, adopting a lower reference enthalpy of fusion for fully crystalline PP, such as 148 J/g[35] instead of 209 J/g, gives a higher value.

## 2.2 Photo-aging

Bending and tensile specimens were photo-aged by an OPM2-502XQ xenon lamp (USHIO Inc., Japan), whose emission spectrum approximates sunlight. Throughout the irradiation, the specimen was held at 70 °C on a house-made hot plate equipped with a temperature controller. The irradiance at the specimen surface was 6.1 $mW/cm^2$ for bending specimens and 5.7 $mW/cm^2$ for tensile specimens, and the irradiation was continued for 12 h. The emission spectrum of the lamp has been reported in our previous study[7]. The characteristic attenuation length of the incident light in the present PP, estimated from the thickness dependence of the transmitted illuminance assuming the Lambert–Beer law, was approximately 4 mm (Appendix A). This is about five times the 0.8 mm specimen thickness used here. Over a 0.8 mm path the attenuation factor is $\exp(-0.8/3.9) \approx 0.81$, i.e., the intensity decreases by only about 19 % across the specimen. Under this irradiation condition, no spontaneous surface cracks were observed prior to mechanical loading. In the following, the irradiated (UV-exposed) face is referred to as the "aged surface" and the opposite, less-aged face as the "reverse surface."

## 2.3 Three-Point Bending Test

Three-point bending tests were performed by a universal testing machine (AGS-X, Shimadzu Corp., Japan) equipped with a 50 N load cell and a three-point bending fixture for plastics. The loading-roller radius was 5 mm and the support-roller radius was 2 mm; the support span was 20 mm. The tests were conducted at a crosshead speed of 1 mm/min at room temperature (24 °C). Two bending configurations were used, and each specimen was tested in only one of them (no specimen was bent in both). In one configuration the aged surface was placed on the tensile side, and in the other the specimen was oriented so that the reverse surface was on the tensile side. Five specimens were tested with the aged surface on the tensile side, and another five with the reverse surface on the tensile side. Therefore, we used ten specimens in total. Prior to testing, the specimen thickness was measured at both ends, outside the support span, and the thickness at the loading point (mid-span) was estimated by linear interpolation. During each test, the side surface of the specimen was recorded with a digital video camera (1280 × 720 pixels). After each test, the tensile-side surface of the specimen was observed with a microscope camera (2592 × 1944 pixels, lens SDS-M, Shodensha Co., Ltd., Japan). Because the support span (20 mm) was much larger than the specimen thickness (0.8 mm), the contribution of shear deformation to the deflection was negligible, and the specimen could be treated within Euler–Bernoulli (pure-bending) theory.

The bending stress at the mid-span tensile surface was calculated as

$$\sigma = \frac{M}{I}\frac{h}{2} = \frac{3lP}{2bh^2}. \tag{1}$$

In Eq. (1), $\sigma$ is stress, $M = Pl/4$ is the bending moment, $I = bh^3/12$ is the second moment of area of the rectangular cross-section, and $h$ is the specimen thickness. $P$ is the load, $l$ is the support span, and $b$ is the specimen width. The corresponding tensile strain at the tensile surface was calculated as

$$\varepsilon = \frac{1}{R}\frac{h}{2} = \frac{6xh}{l^2}. \tag{2}$$

In Eq. (2), $\varepsilon$ is the strain at the tensile surface, $x$ is the mid-span deflection, and $R$ the local radius of curvature of the neutral surface. This definition follows the JIS K7171 and ASTM D790 standards, which assume small-deflection Euler–Bernoulli bending of a simply supported beam under a central point load. The surface strain is evaluated from the mid-span curvature of the theoretical deflection curve; it corresponds to a local radius of curvature $R = l^2/(12x)$ at mid-span.

The validity of the strain given by Eq. (2) is limited by the contact condition at the loading roller. As the deflection increases, the local radius of curvature of the specimen at mid-span decreases, and once the compressive (upper) surface curves as sharply as the loading roller, the specimen begins to conform to the roller. The load is then no longer applied along a single line at mid-span, so that $M = Pl/4$ no longer holds and Eqs. (1) and (2) become invalid.

The local radius of curvature of the neutral surface at mid-span, $R$, is related to the tensile-surface strain by $R = h/(2\varepsilon)$ (From Eq. (2)). The compressive surface, located $h/2$ from the neutral surface, therefore has a radius of $R - h/2$. To keep the contact sufficiently close to a line contact, this radius was required to remain at least three times the loading-roller radius $R_{\mathrm{r}}$ as a safety margin:

$$R - \frac{h}{2} \geq 3R_{\mathrm{r}}. \tag{3}$$

Substituting $R = h/(2\varepsilon)$ into Eq. (3) yields the upper limit of the applicable strain:

$$\varepsilon \leq \frac{h}{6R_{\mathrm{r}} + h}. \tag{4}$$

For $R_r = 5$ mm and a representative specimen thickness of $h = 0.8$ mm, Eq. (4) gives $\varepsilon \leq 0.026$. Note, however, that the tests were continued well beyond this limit (to strains above 0.06).

**2.4 Uniaxial Tensile Test**

Uniaxial tensile tests were carried out by the same universal testing machine with bending test equipped with a 500 N load cell and screw-type flat grips. Specimens were stretched at a constant crosshead speed of 10 mm/min at room temperature, and at least five specimens were tested for each condition. The extensometer (DSES-1000, Shimadzu Corp., Japan) with a 10 mm gauge bar was used to measure the strain. Note that the crosshead speed in the uniaxial tensile test differs from that in the bending test.

### 2.5 Infrared Spectroscopy

The aging state of the aged and reverse surfaces was characterized by infrared (IR) spectroscopy with a Nicolet iS10 FT-IR spectrometer (Thermo Fisher Scientific, USA) operated in the attenuated-total-reflection (ATR) mode. The ATR unit (ATR iTX base, Thermo Fisher Scientific, USA) was equipped with a non-coated diamond internal-reflection element (ATR ID7/ITX, Specac Ltd., UK). Because of the ATR geometry, the spectra probe only a shallow near-surface region; the estimated penetration depth ranges from 0.7 μm at 3000 $cm^{-1}$ to about 2 μm at 1000 $cm^{-1}$. Each spectrum was acquired at a 1 $cm^{-1}$ wavenumber resolution as the average of 64 accumulated scans. The two surfaces and an unaged specimen were measured separately. The degree of photo-aging was quantified by the carbonyl index[13)], defined as the ratio of the integrated absorbance of the carbonyl band centered at 1715 $cm^{-1}$ (assigned to a C=O stretching mode[36)]) to that of a reference band at 2722 $cm^{-1}$ (assigned to a $CH_3$ stretching mode[37)]). Note that no ATR correction was applied. Because the ATR sampling depth depends on the wavenumber, the absolute intensities of bands at different wavenumbers within the same spectrum are not directly comparable; however, since all specimens were measured under identical ATR conditions, the carbonyl indices of different specimens can be compared with one another.

## 3. RESULTS

### 3.1 Evaluation of the Aging Gradient by FT-IR

Figure 1 shows the IR spectra of the aged surface, the reverse surface, and the unaged specimen. IR spectra were normalized by the area of the peak at 2722 $cm^{-1}$. A pronounced carbonyl band centered at about 1715 $cm^{-1}$ was observed for the aged surface, a weaker band for the reverse surface, and essentially no band for the unaged specimen. To calculate the carbonyl index, as shown in Fig. 2, the spectrum in the 1800-1600 $cm^{-1}$ region was decomposed into two components centered at 1715 and 1640 $cm^{-1}$. The bands at 1640 $cm^{-1}$ are attributed to C=C stretching[38)]. The carbonyl index decreased in the order of the aged surface, the reverse surface, and the unaged specimen, with values of 6.17 for the aged surface, 4.56 for the reverse surface, and a negligible value (~0.00) for the unaged specimen. This result confirms the existence of an aging gradient across the specimen thickness, with the aged surface being more strongly photo-oxidized than the reverse surface. To place these values in context, Rabello and White used the same carbonyl-index definition (normalized to the 2722 $cm^{-1}$ band) but calculated it from spectra measured in transmission mode. Although a direct comparison is not possible because the measurement methods differ, their reported indices rose from a few units at early UV exposure to about 8–12 after 12 weeks and 20–33 after 24 weeks[13)]. The present values (4.56 and 6.17) therefore correspond to a mild stage of photo-oxidation, well below the heavily degraded regime.

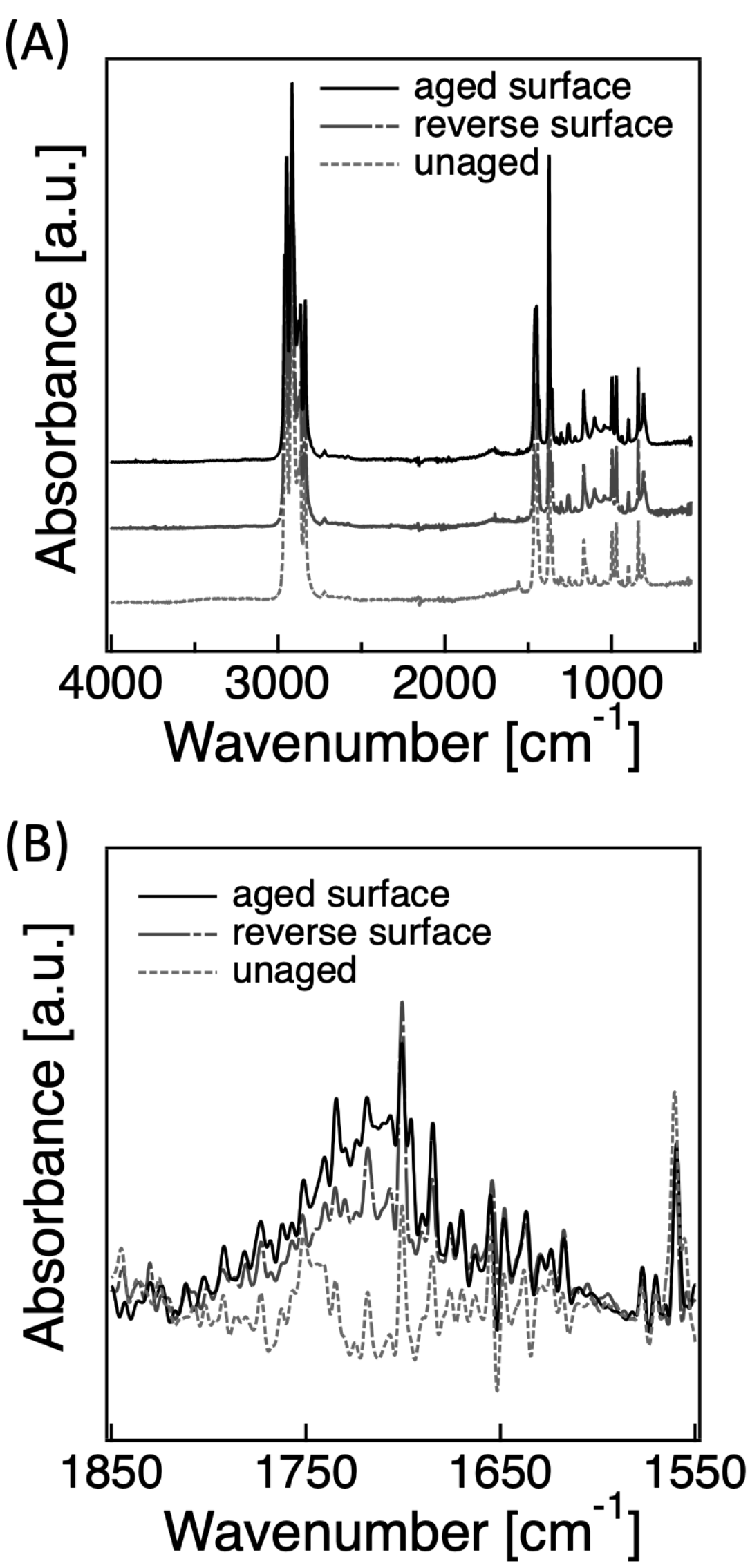

Figure 1 IR spectra of the aged surface, the reverse surface, and the unaged specimen in the wavenumber ranges of (A) 4000–500 $cm^{-1}$ and (B) 1850–1550 $cm^{-1}$, normalized by the area of the 2722 $cm^{-1}$ band. No ATR correction is applied.

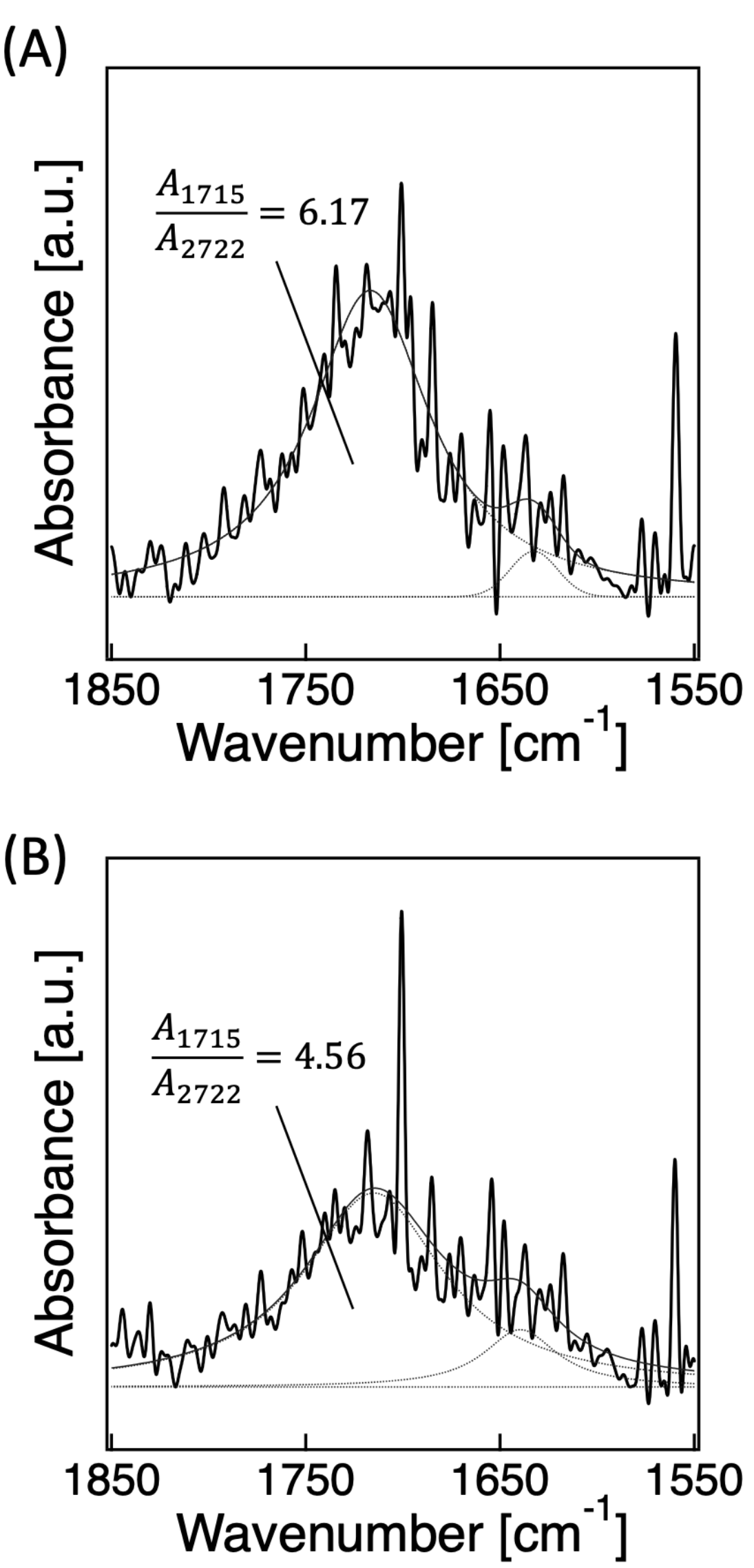

Figure 2 IR spectra for 1850–1550 $cm^{-1}$ and peak fitting results of (A) the aged surface and (B) the reverse surface. The IR spectra are normalized by the area of the 2722 $cm^{-1}$ band and corrected for the baseline. No ATR correction is applied. $A_i$ represents integrated absorbance of the peak centered at $i$ $cm^{-1}$. $A_{1715}/A_{2722}$, shown in the figure, is the carbonyl index.

### 3.2 Ductile-to-Brittle Transition under Uniaxial Tension

Before turning to the bending experiments, we first examine what information a conventional uniaxial tensile test provides for the present photo-aged PP. Tensile testing gives a stress–strain curve up to a break point.

Figure 3 shows stress–strain curves of the unaged specimen and of the photo-aged specimens. For each condition, a single representative curve is shown, although at least five specimens were tested (Section 2.4). The unaged PP exhibited a typical ductile stress–strain response. After an initial quasi-linear elastic region, the nominal stress reached a yield point (the yield stress and strain were 30 MPa and 0.2, respectively). Beyond the yield point the stress decreased and a neck was formed in the gauge section. The neck then propagated at an approximately constant stress at about 26 MPa, producing the extended plateau between the strain range about 0.6-3. At large strains (above about 3) the stress rose again owing to strain hardening of the drawn material, and the specimen finally fractured. This sequence—elastic deformation, yielding, neck propagation, and strain hardening up to ductile failure—is characteristic of semi-crystalline PP in the unaged state[39,40)]. After 12 h of photo-aging, the tensile behavior of PP changed markedly, shifting from ductile to brittle[8,9,14)]. Whereas the unaged specimens deformed in a fully ductile manner, the photo-aged specimens exhibited a pronounced loss of ductility: the elongation at break decreased drastically, and neither yielding nor neck propagation was observed. In parallel, the initial slope of the stress–strain curve, which corresponds to the Young's modulus, increased due to the chemi-crystallization.

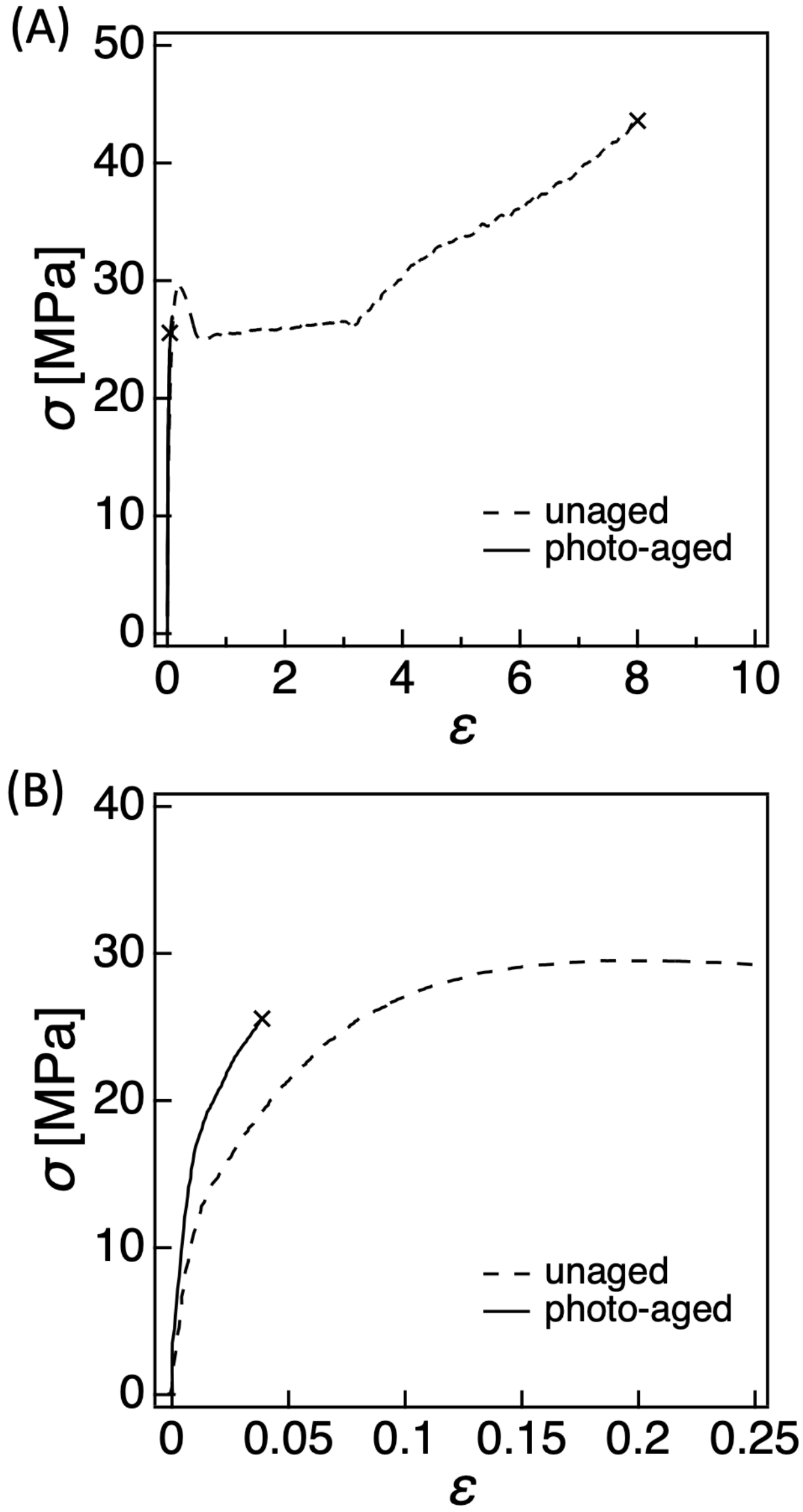

Figure 3 Stress–strain curves obtained from uniaxial tensile tests of the unaged specimen and of the photo-aged specimen in the strain ranges of (A) 0–10 and (B) 0–0.25. The points at which the specimens break are highlighted by cross marks.

### 3.3 Stress–Strain Curves under the Two Loading Configurations

We now turn to the three-point bending experiments and examine what information they provide for the same photo-aged PP. Unlike uniaxial tension, which loads the entire cross-section and cannot resolve the aged and reverse surfaces separately, three-point bending subjects a single chosen face to the maximum tensile stress; placing either surface on the tensile side thus yields distinct stress–strain curves.

Figure 4 shows stress–strain curves of the unaged specimen and of the photo-aged specimens loaded with the aged or the reverse surface on the tensile side. Beyond a strain of approximately 0.026 (Eq. (4)), the loading roller no longer contacted the specimen along a single line; the response in this region should therefore be interpreted with caution. Both photo-aged configurations exhibited higher stress than the unaged specimen even at small strains, indicating that photo-aging increased the bending modulus of the specimen[8]. The same increase is observed in Fig. 3. This increase refers to the difference between the photo-aged and unaged specimens and should not be confused with the comparison between the two photo-aged configurations, which is discussed separately below.

Between the two photo-aged configurations, no appreciable difference was observed in the small-strain region. However, a clear difference between the configurations appeared at larger strains. When the aged surface was placed on the tensile side, the stress–strain curves of three specimens exhibited several discrete stress drops, while the other two specimens completely fractured at the first drop. This specimen-to-specimen variability—several stress drops in some specimens and a single catastrophic fracture in others—reflects the defect-controlled randomness intrinsic to photo-aged specimens; nevertheless, the three specimens that did not fracture catastrophically shared a closely similar overall curve shape and differed mainly in the positions of the stress drops. As shown in Section 3.4, these stress drops originate from crack formation on the tensile-side surface. In contrast, when the reverse surface was placed on the tensile side, no clear stress drops were observed; at most, mere kinks that did not develop into full drops appeared in the high-strain region. In addition, the maximum stress attained by the specimens that exhibited stress drops was lower than that of the specimens loaded with the reverse surface on the tensile side. The stress–strain curves and stress drops for all tests are shown in Figure A2 in Appendix B.

We briefly comment on the applicable range of the bending strain, which Eq. (4) limits to $\varepsilon \leq 0.026$ for the present specimens. The factor of three in Eq. (3) used to define this threshold was a conservative safety margin rather than a strict physical limit, adopted because crack formation on the tensile surface (the stress drops described above) locally concentrates the curvature and reduces the actual radius of curvature below the value estimated from the overall deflection. Beyond this strain the calculated stress and strain become quantitatively unreliable—indeed, they are already approximate once a surface crack has formed—but the discrete stress drops and the surface cracks can still be counted reliably. The crack/stress-drop comparison presented below, and the micrographs taken after bending to strains above 0.06, therefore rely only on counting these discrete events and not on the absolute stress or strain values.

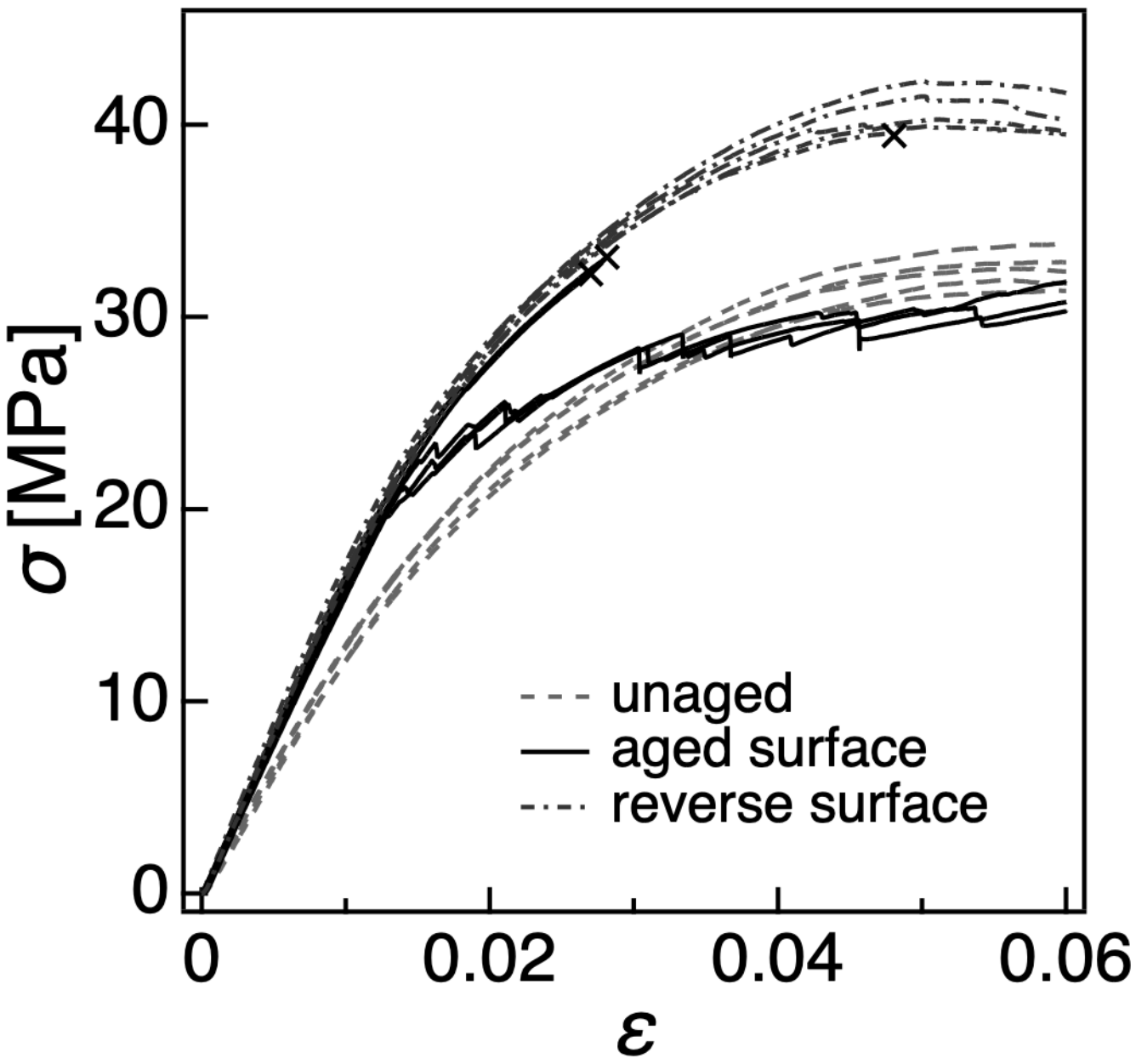

Figure 4 Stress–strain curves obtained from three-point bending tests of the unaged specimen and of the photo-aged specimens loaded with the aged surface or the reverse surface on the tensile side. The points at which the specimens break are highlighted by cross marks.

### 3.4 Correspondence between Stress Drops and Surface Crack Formation

Figure 5 shows (A) the stress–strain curve and (B–D) side-view images of the corresponding specimen, captured during a bending test with the aged surface on the tensile side over the strain range 0.012–0.018; the contrast of (B–D) has been adjusted to enhance visibility. Before each stress drop the tensile surface appeared smooth, and at each stress drop a crack appeared at the tensile-side surface simultaneously with the load decrease; that is, every newly formed crack coincided with a stress drop. The first crack was identified at a bending strain of approximately 0.0152. The images also showed clearly that each crack initiated from the aged surface and arrested within the specimen before reaching the opposite (compressive) side, propagating only to a finite depth from the surface. The side-view images could not, however, reveal whether an existing crack subsequently advanced deeper; therefore, when a crack advanced in a stepwise manner, we could not determine whether the later advance steps were each accompanied by a stress drop. The stress drops varied in magnitude, but all were counted, giving 15 in total for this specimen.

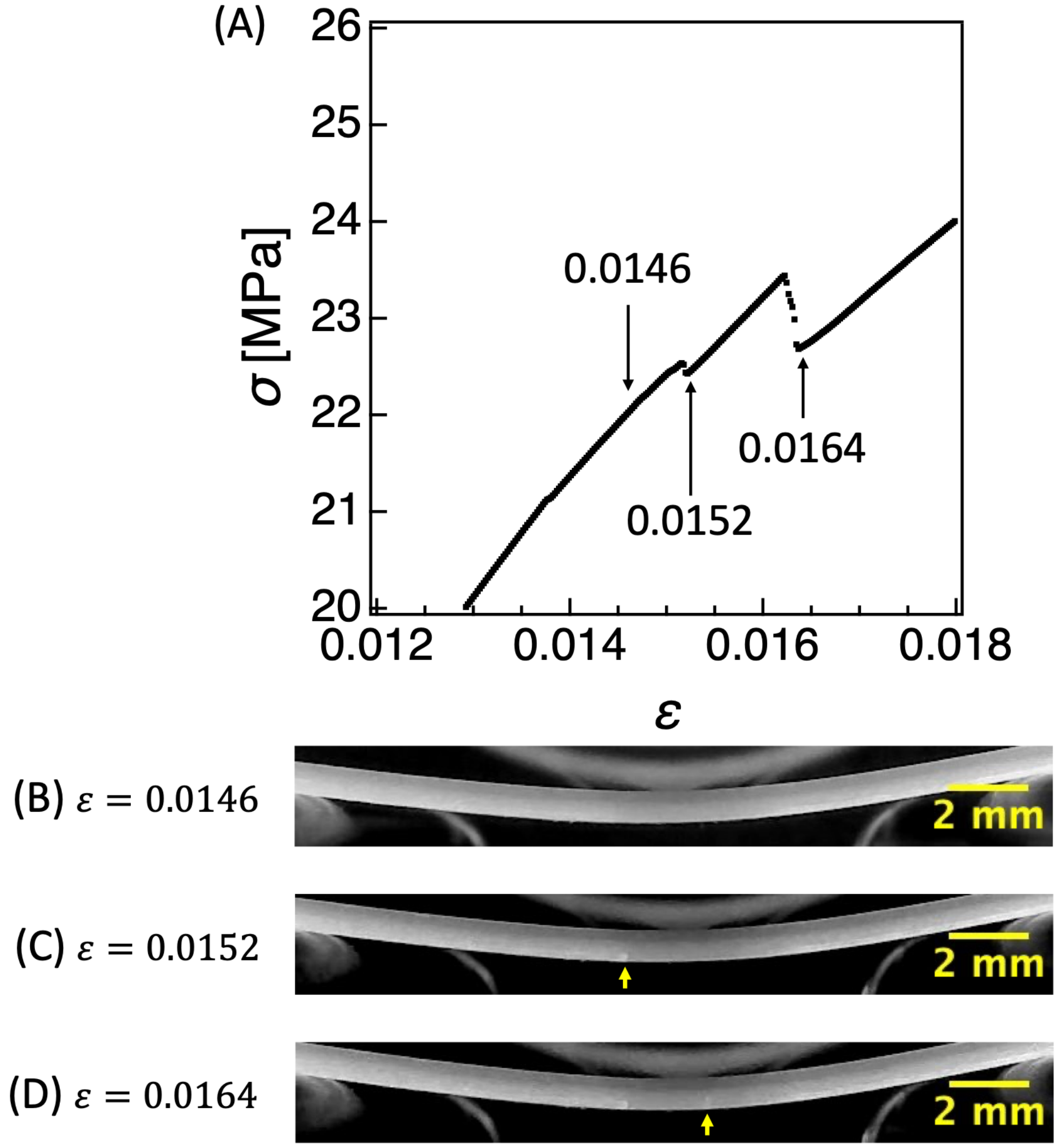

Figure 5 (A) Stress–strain curve and (B–D) side-view images of the corresponding photo-aged specimen during a three-point bending test with the aged surface on the tensile side, in the strain range of 0.012–0.018. The strains at which the side-view images (B)–(D) were captured are indicated on the curve in (A) by their strain values. Newly formed cracks are marked by arrows in (C) and (D).

Figure 6 shows optical micrographs of the tensile-side surface of photo-aged specimens after the bending tests, for the two loading configurations. The contrast of Fig. 6 (A–B) has been adjusted to enhance visibility. Before testing, both surfaces were smooth and exhibited no visible cracks. After bending to a strain larger than 0.06, when the aged surface was on the tensile side, ten nearly straight cracks formed perpendicular to the tensile direction across the surface. These 10 cracks were fewer than the 15 stress drops observed in the corresponding stress–strain curve. Although the two counts do not match exactly, these observations are qualitatively consistent with the stress–strain curves in Fig. 5 and confirm that the multiple stress drops observed in the aged-surface configuration correspond to the formation of multiple surface cracks. The *in-situ* observation showed that the formation of each new crack at a previously uncracked location coincided with a stress drop, so that at least 10 of the 15 counted stress drops corresponded to the 10 surface cracks. The *in-situ* observation also revealed the spatial sequence of cracking. The first crack formed in the most highly stressed region near the center of the specimen, which corresponds approximately to the loading-roller position. Subsequent cracks then appeared in the regions between the first crack and the supports. These cracks did not necessarily form in sequence from the center toward the supports; rather, they developed at spaced intervals, separated from one another. In addition, the crack spacing increased with distance from the center of the loading span; in other words, the cracks were not randomly distributed. In contrast, when the reverse surface was on the tensile side, only several cracks appeared near the specimen edges, consistent with the absence of clear stress drops in the corresponding stress–strain curve (Fig. 4). Micrographs for all specimens and a comparison of the numbers of stress drops and surface cracks are provided in Figure A3 (Appendix C) and Table A1 (Appendix D), respectively.

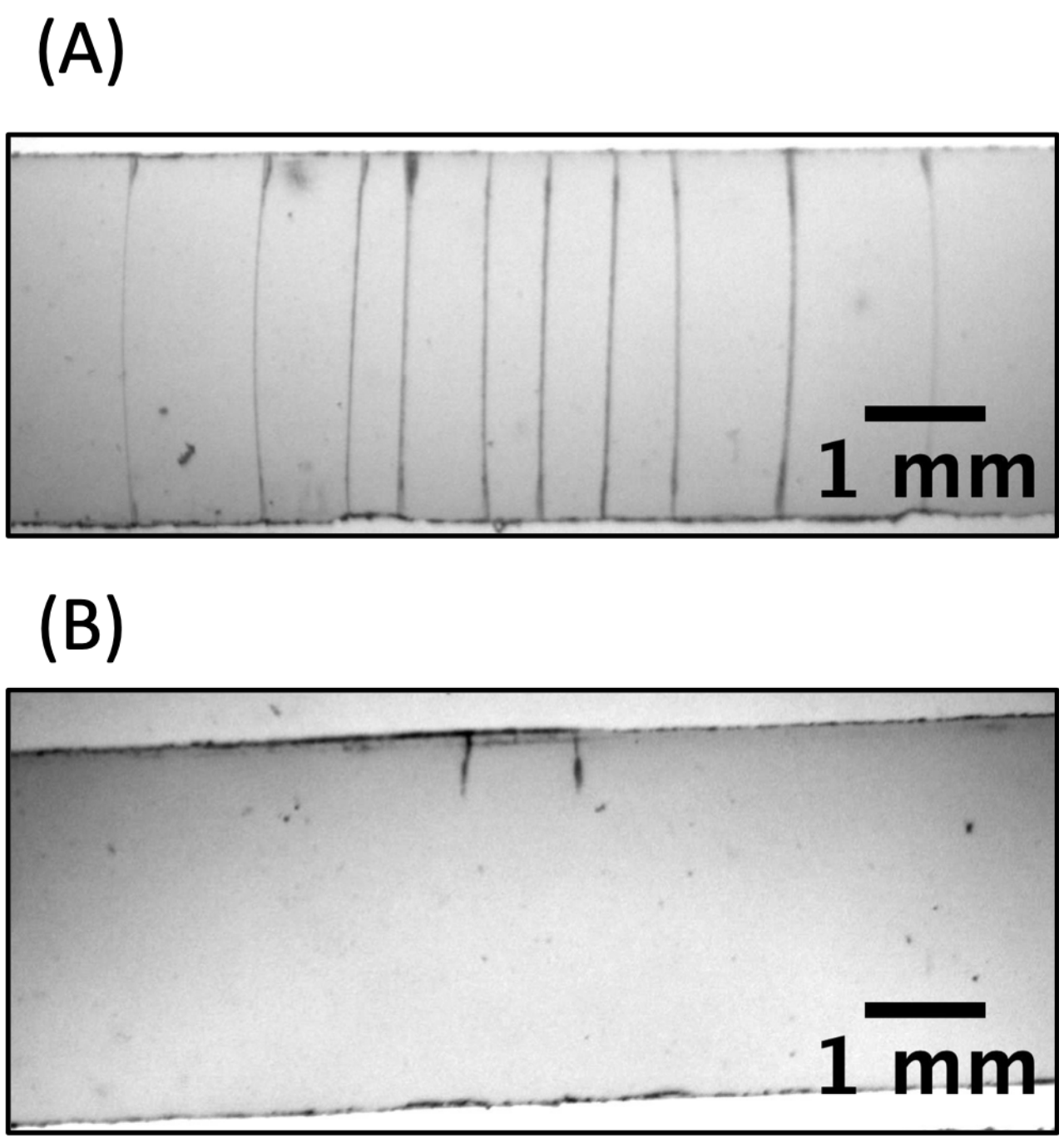

Figure 6 Optical micrographs of the tensile-side surface of photo-aged specimens after three-point bending to a strain larger than 0.06; (A) the aged surface on the tensile side and (B) the reverse surface on the tensile side. The same specimen as that in Fig. 5 is shown in (A). The image center corresponds approximately to the loading-roller position.

## 4. DISCUSSION

The results above show that the stress–strain curve obtained from three-point bending reflects the aging gradient of the photo-aged PP specimen through the appearance of stress drops associated with surface crack formation. Because the specimen is structurally non-uniform across its thickness, it is self-evident that flipping it changes the response; the non-trivial point, discussed below, is that bending renders the surface embrittlement directly visible as discrete stress drops and surface cracks, and that these appear only when the embrittled surface is placed on the tensile side.

Photo-oxidation of PP induces chain scission, which lowers the molecular weight and reduces the local stress at break of the surface layer. Chain scission also enhances the mobility of the shortened chains and promotes chemi-crystallization, in which these fragments partially reorganize into new crystals. Chemi-crystallization increases the local density and is accompanied by a volume shrinkage of the aged layer[13]. Because the degree of aging decreases with depth, both the fracture toughness and the amount of shrinkage are distributed across the thickness[11]. In principle, such a gradient could affect the bending response in two ways: through a gradient of the elastic modulus (and the associated shrinkage) and through a gradient of the fracture toughness. Note that both the UV intensity and oxygen diffusion govern the aging distribution. Because the UV intensity differs by only approximately 19 % between the aged and reverse surfaces (as mentioned in Section 2.2), the through-thickness aging distribution—which extends far enough into the specimen to stop crack propagation—is attributed mainly to oxygen diffusion. In other words, the overall shape of this distribution, in which a brittle surface layer several hundred micrometers thick grades into the ductile interior, is controlled by oxygen diffusion. The degree of aging, however, differs slightly between the two surfaces (as mentioned in Section 3.1) correlating with the small difference in UV intensity, and it is this slight difference that determined whether surface cracks formed during the bending test, giving rise to the contrasting crack-initiation behavior of the aged and reverse surfaces.

In the small-strain region during the bending test, the stress–strain curves for the two configurations almost coincide (Fig. 4). This indicates that the modulus gradient and the shrinkage associated with the aged layer are not large enough to be reflected in the overall bending response. This near coincidence is, in fact, what a modulus contrast alone would predict (Appendix E). When the specimen is modeled as a two-layer beam with a stepwise modulus change at the mid-plane, the neutral surface lies at

$$\xi = \frac{h}{2} + \frac{h}{8}\frac{\Delta E}{\bar{E}}, \tag{5}$$

where $\xi$ is position of the neutral surface (zero-strain plane) measured from the upper surface. $\bar{E}$ and $\Delta E$ are mean modulus and modulus difference between the layers, respectively. The

moment–curvature relation is

$$M = \frac{\bar{E}bh^3}{12}\left[1 - \frac{3}{16}\left(\frac{\Delta E}{\bar{E}}\right)^2\right]\frac{1}{R}. \tag{6}$$

Because the $\Delta E$ term enters only as its square, the relation is independent of which face is placed upward; that is, the load and the deflection take the same values in the two orientations. In the present analysis, the stress and strain are calculated under the assumption that the neutral surface always lies at the mid-plane (Eqs. (1) and (2)); consequently, this calculation scheme yields no difference in the surface strain or stress between the two orientations, and the apparent Young's modulus is identical. Although this is only a simple-model calculation, it agrees well with the experimental results and is considered to be one of the reasons why the two configurations exhibit no appreciable difference in the small-strain region.

In contrast, the two configurations differ markedly at larger strains, where crack formation begins to govern the response. When the aged surface is placed on the tensile side, the largest tensile stress acts on the most severely aged and embrittled region. As the bending strain increases, the tensile stress at the surface eventually exceeds the local stress at break of the aged layer, and a crack forms. This local stress at break of the embrittled surface is also reflected in a conventional uniaxial tensile test. Indeed, as shown in Section 3.2 (Fig. 3), the photo-aged specimens broke at a much smaller strain than the unaged specimens—that is, their elongation at break was drastically reduced—consistent with this picture. The crack then propagates into the interior under the stress concentration at its tip, but the fracture toughness increases with depth as the aging becomes weaker; the crack is therefore stopped within the interior[11,13,21]. When a crack forms and is stopped, the layer that had been carrying tensile load over the crack depth no longer contributes, and the load supported by the specimen decreases, which appears as a stress drop in the stress–strain curve. After the drop, the load is redistributed and increases again until the tensile stress reaches the local stress at break at another location, producing the next crack and the next stress drop. This repeated process accounts for the series of stress drops and the multiple nearly straight cracks observed on the aged surface. This crack-stop picture is directly supported by the side-view observation in Fig. 5, where each crack is seen to initiate from the aged surface and stop within the interior, rather than propagating across the entire thickness.

The multiple-cracking behavior observed here closely resembles that reported for thin, hard, brittle coatings on ductile metallic substrates[29,41)]. Photo-aging generates a thin embrittled layer on an otherwise ductile bulk, which is mechanically analogous to a brittle coating bonded to a tough substrate.

The cracks were not randomly distributed but tended toward a regular spacing (Fig. 6A), again in agreement with the coating literature[30,41)]. When a crack forms, the tensile stress recovers from zero at the crack face to the far-field value only over a characteristic distance on either side of the crack. Within this recovery zone the stress cannot reach the level required for fracture, and the nucleation of new cracks is suppressed nearby. This stress-transfer process is analogous to the shear-lag model used to describe load transfer to short fibers in composites, and it underlies some fracture models[32,41)]. Because no further cracks can form within a certain length of an existing crack, the resulting crack pattern is more regularly spaced than would be expected from a purely random flaw population, consistent with our observation that the cracks were not randomly distributed.

If any difference from the nearly constant crack spacing reported in earlier studies is to be pointed out, it is that the crack spacing increased systematically with distance from the center of the loading span. This follows directly from the difference between three- and four-point bending. The coating analyses of Gille[32)] and Ramsey, Chandler and Page[41)] were formulated for the constant-moment region between the inner rollers of a four-point fixture, where the surface tensile stress is spatially uniform and the cracks are therefore expected to be evenly spaced. In three-point bending, by contrast, the bending moment is maximal directly beneath the central loading point and decreases linearly toward the supports, producing a tensile-stress gradient along the specimen surface. Therefore, the cracks are most closely spaced beneath the central loader, where the stress is highest, and become progressively more widely spaced toward the supports, where the stress falls below the threshold for fracture. Accordingly, while the statistical framework of Gille[32)] is geometry-independent in principle, it assumes the uniform-stress field of four-point bending and would need to incorporate the position-dependent stress to describe the three-point results quantitatively.

A slight discrepancy was found between the numbers of cracks and stress drops. This may be attributed to two factors. First, a single crack may not propagate to completion in one step but instead advance gradually in two or three stages, so that one crack can give rise to several stress drops. Second, when cracks form at nearly the same time, or when a crack propagates slowly, the resulting stress drops may overlap and become difficult or impossible to resolve individually. In the present system, the stress–strain curves exhibited a range of features, from large stress drops to mere kinks that did not develop into full drops, and all of these were counted as stress drops. However, since large stress drops are expected to correspond to large crack advances, an analysis focused on the large stress drops may in fact be the more important one.

When the reverse surface is placed on the tensile side, the surface subjected to the largest tensile stress is much less aged and retains a larger local stress at break. Within the strain range examined, the tensile stress only rarely exceeds the local stress at break of this surface, so that far fewer cracks form and no clear stress drops appear, in contrast to the aged-surface configuration. The small cracks observed near the edges (Fig. 6B) are attributed to the local stress concentration at the free edges of the specimen arising from the specimen geometry. The contrast between the two configurations therefore demonstrates that the bending test selectively reflects the fracture toughness of the surface placed on the tensile side, that is, the surface aging state. It should be noted that unaged semicrystalline PP fails in a ductile manner, so that crack growth is preceded by substantial plastic deformation around the crack tip[40]; embrittlement by photo-oxidation lowers this plastic dissipation and the fracture toughness, which is why discrete surface cracks appear only on the aged side. Whether a crack then propagates is, in the brittle limit, governed by the local stress at break measured in a fracture test rather than by the uniaxial tensile response. Schoolenberg and Vink reported a fair correlation between the carbonyl index and mechanical embrittlement in photo-degraded PP[26]. It should be emphasized, however, that the carbonyl index is a chemical measure and does not translate directly into mechanical embrittlement: Fayolle et al. found that PP can embrittle before the end of the carbonyl induction period and at a very low degree of chain scission, so that mechanical embrittlement is far more sensitive than the carbonyl index[1]. Also, this transition is not simply monotonic with exposure: Rabello and White reported that at longer exposure a ductile band re-forms beneath the embrittled surface layer and stops the surface cracks, so that the apparent tensile properties can partially recover[13].

A particular advantage of the bending geometry for photo-aged materials is that the largest tensile stress is generated exactly at the surface, where the aging is most severe. Uniaxial tensile testing, by contrast, assumes a nominally uniform deformation across the cross-section and therefore reflects the spatially averaged properties. The present results show that three-point bending, combined with *in-situ* observation, can detect crack formation that reflects the surface aging state of mildly photo-aged specimens that do not yet exhibit spontaneous cracks. Compared with local probes such as nanoindentation, this approach requires no specialized instrumentation and reveals macroscopic crack formation—several cracks from a single specimen—directly correlated with the load response through *in-situ* observation. Compared with uniaxial tension, it allows successive surface cracks to be observed without the specimen breaking apart, and it preferentially extracts the surface response rather than the bulk-average behavior. In addition, despite the large apparent deflection, the strain experienced at the surface remains very small, so that the small-strain region can be examined in detail. Because photo-aging induces a pronounced ductile-to-brittle transition, three-point bending is particularly well suited to polypropylene. The response remains stable even after the specimen has cracked, which reflects the robustness of the measurement. The test is also experimentally convenient: uniaxial tensile testing requires the specimen to be gripped at both ends, which can damage or break it during clamping, whereas in bending the specimen is simply rested on the supports and pressed by the loading roller, causing little damage and further improving the stability of the data. The ability to acquire data on surface-crack formation that cannot be detected by uniaxial tension is therefore highly valuable.

Looking beyond the present results, we note two further directions suggested by our findings. The first, which appears experimentally feasible, is that the crack formation observed in the present bending tests may carry information about the spontaneous surface cracking that occurs during photo-aging. As photo-aging progresses, internal shrinkage stresses develop within the specimen while the material simultaneously becomes more prone to fracture. Because the aging distribution is non-uniform across the thickness, the shrinkage-stress distribution is also non-uniform, and the aged surface is effectively subjected to a stress state similar to that of the tensile side in bending deformation. In the bending test, by contrast, the stress is supplied externally, and cracks form once the tensile stress reaches the local stress at break, which sets the fracture threshold of the aged surface. Because this threshold decreases with the progress of photo-aging, performing bending tests on specimens with different degrees of aging and tracking the change in the threshold may allow the degree of aging at which spontaneous cracks form to be estimated.

The second, more exploratory direction starts from the correspondence between each stress drop and a surface crack, which suggests that quantitative information may be extracted from the magnitude of the stress drops. If a stress drop reflects the loss of the load carried by the cracked layer, its magnitude is expected to be related to the depth to which the crack propagates before stopping, which in turn is governed by the depth profile of the fracture toughness. We note, however, that crack stopping reflects the combined effects of the plastic dissipation, ductile deformation, and stress redistribution; isolating the toughness profile quantitatively is therefore very difficult, and the present analysis remains qualitative on this point. A quantitative analysis relating the stress-drop magnitude, the crack depth, and the fracture toughness is beyond the scope of the present study and remains an important subject for future work.

## 5. CONCLUSION

Three-point bending tests were applied to mildly photo-aged PP specimens that had an embrittlement gradient across their thickness but had not developed spontaneous surface cracks before loading. When the aged surface was placed on the tensile side, the stress–strain curves exhibited several discrete stress drops, and *in-situ* side-view observation confirmed that the formation of each new surface crack coincided with a stress drop. When the less-aged reverse surface was placed on the tensile side, no clear stress drops were observed and only a few small cracks formed near the edges. This contrast between the two loading configurations shows that the degree of surface embrittlement, and hence its variation across the thickness, is directly reflected in the stress–strain curve of a three-point bending test. Bending thus provides an effective probe of surface embrittlement and its through-thickness gradient that uniaxial tensile testing, which reflects the spatially averaged response, cannot easily provide.

## APPENDIX

## A. Light-Attenuation (Characteristic) Length of the PP Specimens

### A.1 Specimen Preparation

The specimens used for this evaluation were prepared by the same procedure as described in Section 2.1, except that the thickness of the aluminum spacer was varied. Spacers of 0.1, 0.3, 0.5, and 1.0 mm (150 mm × 150 mm, with a 100 mm × 100 mm central window) were used in place of the 1 mm spacer employed for the specimens in the main text. After hot-pressing at 200 °C and 15 MPa for 15 min and quenching in ice water at 0 °C, the sheets were cut into circular specimens 10 mm in diameter. After annealing at 140 °C for 3 h, the measured thicknesses of the specimens were 0.102, 0.287, 0.452, and 0.812 mm.

### A.2 Illuminance Measurement

Each specimen was placed normal to the collimated UV light described in Section 2.2, and the transmitted illuminance $I$ at 365 nm was measured by a UIT-250 cumulative UV illuminance meter (USHIO Inc., Japan) and a UVD-S365 photodetector (USHIO Inc., Japan), with the specimen placed on top of the detector. The illuminance transmitted in the absence of any specimen, $\Phi_0 = 5.49\ \mathrm{mW/cm^2}$, was recorded as a reference. All measurements were carried out at room temperature.

### A.3 Analysis and Results

Assuming the Lambert–Beer law, the transmitted illuminance decays with specimen thickness $z$ as

$$\Phi(z) = \Phi_0 \exp\left(-\frac{z}{z_0}\right), \quad \text{(A1)}$$

where $z_0$ is the light-attenuation (characteristic) length and $\Phi_0$ the incident illuminance. The characteristic attenuation length is the depth over which the transmitted intensity falls to $1/e$ of its incident value. Taking the logarithm gives a linear relation, so that $z_0$ is obtained as the reciprocal of the slope of a linear fit of $\ln \Phi$ against $z$.

Figure A1 shows the relation between the sample thickness and transmitted light intensity. A least-squares fit to all four specimens gave $\ln(\Phi\ /\mathrm{mWcm^{-2}}]) = (1.692 \pm 0.018) - (0.26 \pm 0.03)(z\ /\mathrm{mm})$, corresponding to $z_0 \approx 3.9\ \mathrm{mm}$ and an extrapolated incident illuminance $\Phi_0 = 5.43\ \mathrm{mW/cm^2}$. This characteristic length of about 4 mm is roughly five times larger than the 0.8 mm specimen thickness used in the main text. Over a 0.8 mm path, the attenuation factor is $\exp(-0.8/3.9) \approx 0.81$, i.e., the intensity decreases by only about 19 % across the specimen. Thus, the difference in crack initiation between the aged and reverse surfaces arises from the slight difference in aging degree between them, which in turn reflects this small difference in UV intensity, whereas the aging distribution formed across the thickness is governed by $O_2$ diffusion.

Several points concerning the reliability of this estimate are worth noting. First, though all specimens were prepared by the same procedure as in the main text and differed only in spacer thickness, a different specimen thickness leads to a somewhat different cooling history and hence to changes in the higher-order structure (crystallinity, spherulite size, etc.). The effect of such structural changes is gradual and is not expected to alter the attenuation coefficient by an order of magnitude over the present thickness range. Second, for thicker specimens, internal (multiple) scattering causes the data to deviate from the simple straight line, with $\ln \Phi$ tending to fall below the linear extrapolation. To assess the influence of this effect, the fit was repeated using only the three thinnest specimens, for which scattering is least significant; this gave $z_0 \approx 6.3\ \mathrm{mm}$. The two estimates ($\approx 4\ \mathrm{mm}$ and $\approx 6\ \mathrm{mm}$) remain within the same order of magnitude. The characteristic length is therefore robustly of the order of several millimeters, and in all cases substantially larger than the 0.8 mm specimen thickness. Finally, the incident illuminance $\Phi_0$ obtained by extrapolation ($5.43\ \mathrm{mW/cm^2}$) is slightly lower than the value measured without a specimen ($5.49\ \mathrm{mW/cm^2}$). The difference reflects the magnitude of light lost to surface reflection and scattering.

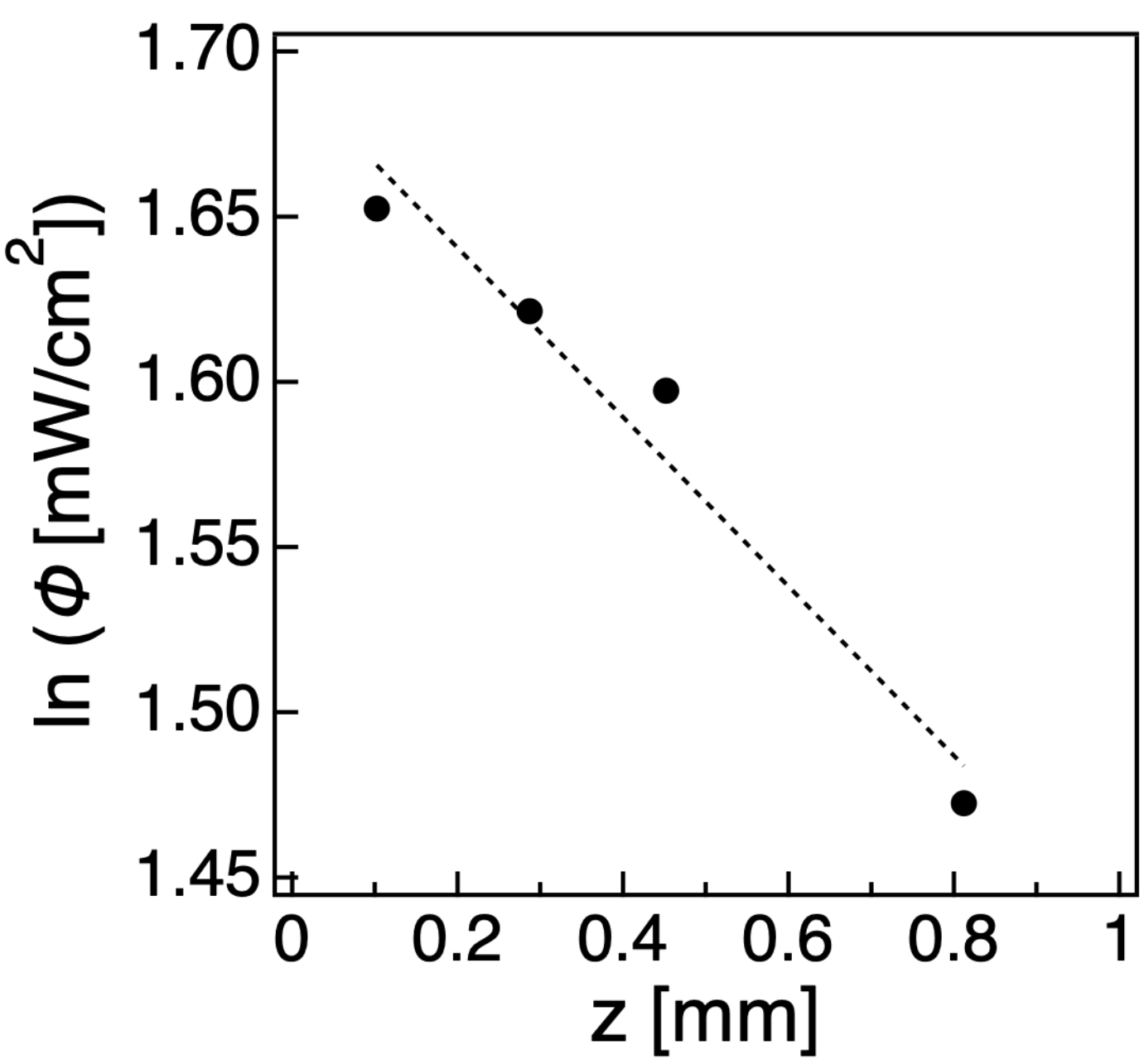

Figure A1 Relation between the specimen thickness and the logarithm of the transmitted light intensity. The dotted line shows the result of fitting to the Lambert–Beer law.

## B. Stress–Strain Curves and Stress Drops for All Bending Tests

Figure A2 shows the stress–strain curves obtained from all five three-point bending tests performed for each loading configuration, together with the stress drops identified along each curve. Each curve is labeled with a specimen identifier (a1–a5 for the aged-surface configuration and b1–b5 for the reverse-surface configuration); the same identifiers correspond to the same specimens in Figure A3 and Table A1.

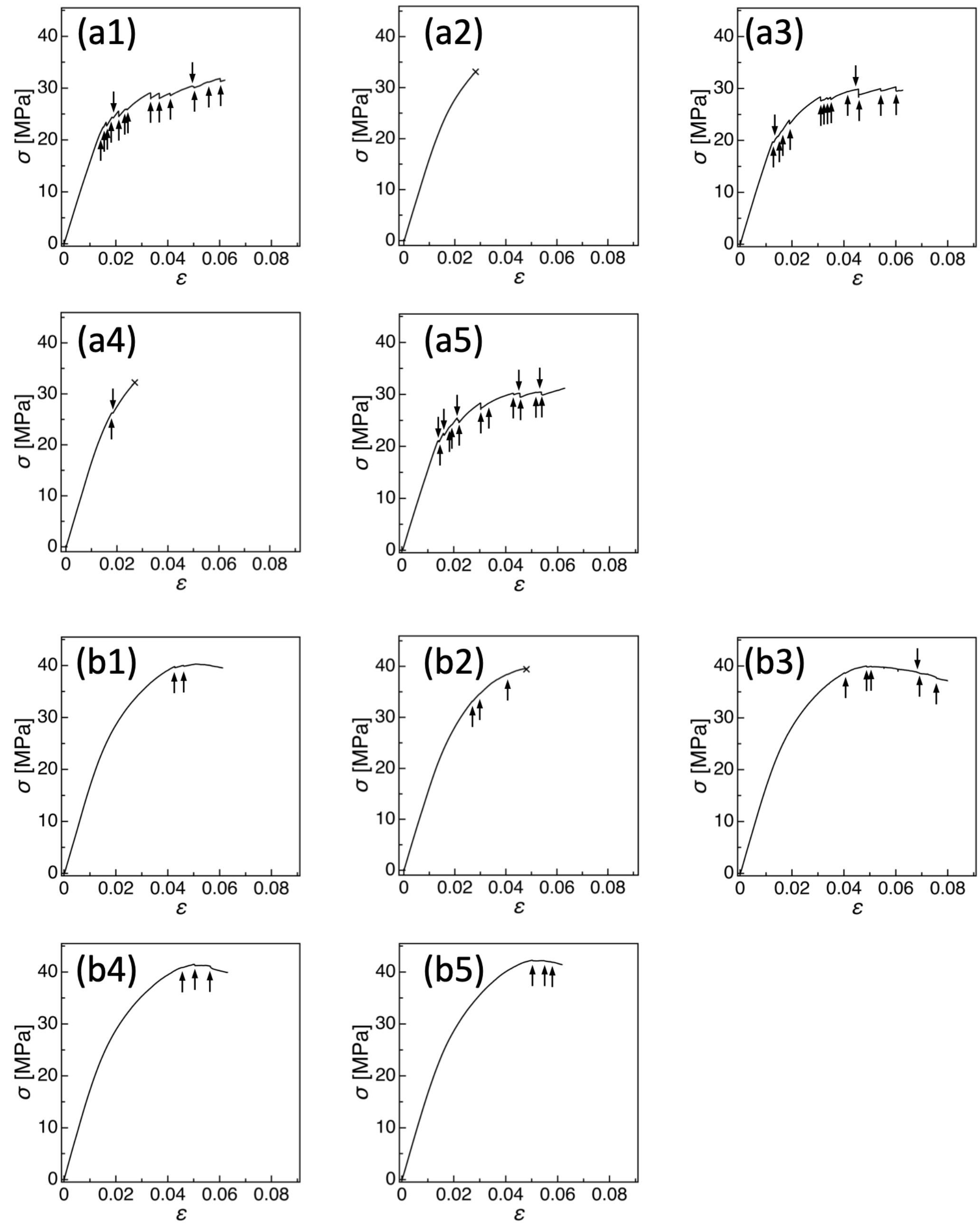

Figure A2 Stress–strain curves obtained from all bending tests, for the configurations with (a) the aged surface on the tensile side and (b) the reverse surface on the tensile side. Each curve is labeled with a specimen identifier (a1–a5 for the aged-surface configuration and b1–b5 for the reverse-surface configuration); the same identifiers are used in Figure A3 and Table A1 for the same specimens. The stress drops identified in each curve are indicated by black arrows. The points at which the specimens break are highlighted by cross marks.

## C. Optical Micrographs of the Tensile-Side Surface for All Specimens

Figure A3 shows optical micrographs of the tensile-side surface of all photo-aged specimens after three-point bending to a strain larger than 0.06, for the two loading configurations. Each micrograph is labeled with the specimen identifier (a1–a5 and b1–b5), corresponding to the same specimens shown in Figure A2 and Table A1. The micrographs shown in Fig. 6 (A) and (B) of the main text are the same as (a1) and (b1) in Figure A3, respectively, and are included here for reference.

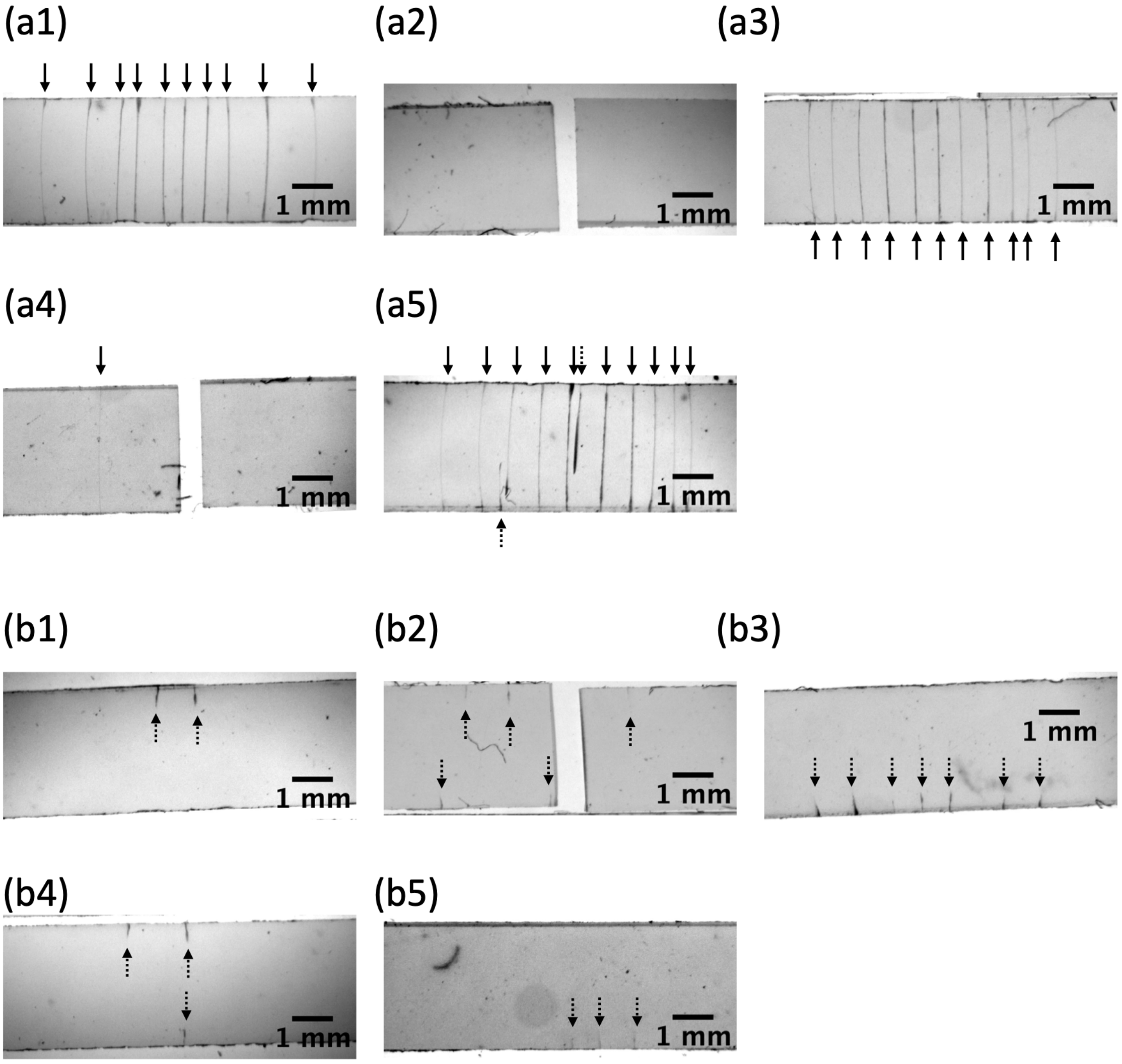

Figure A3 Optical micrographs of the tensile-side surface of all photo-aged specimens after three-point bending to a strain larger than 0.06; (a) the aged surface on the tensile side and (b) the reverse surface on the tensile side. Each micrograph is labeled with the specimen

identifier (a1–a5 and b1–b5), which corresponds to the same specimens shown in Figure A2 and Table A1. The image center corresponds approximately to the loading-roller position. The straight cracks across the surface and small cracks near the specimen edges are indicated by solid and dotted arrows, respectively.

## D. Comparison of the Number of Stress Drops and Surface Cracks for All Specimens

For each specimen tested with either the aged surface (a1–a5) or the reverse surface (b1–b5) on the tensile side, the number of stress drops identified in the stress–strain curve (Figure A2) was compared with the number of surface cracks counted in the corresponding optical micrograph (Figure A3). In this counting, a "stress drop" refers not only to a clear stress drop but also to a smaller kink-like feature that did not develop into a full drop. The comparison for all specimens is summarized in Table A1, in which the specimen identifiers (a1–a5 and b1–b5) correspond to those shown in Figures A2 and A3.

Table A1 Number of cracks and stress drops counted for each specimen tested with the aged surface on the tensile side (a1–a5) and the reverse surface on the tensile side (b1–b5). The specimen identifiers correspond to those used in Figures A2 and A3.

| Specimen | Tensile-side configuration | Straight cracks | Edge cracks | Stress drops | Note |
|---|---|---|---|---|---|
| a1 | Aged surface | 10 | 0 | 15 | |
| a2 | Aged surface | 0 | 0 | 0 | Fractured |
| a3 | Aged surface | 11 | 0 | 14 | |
| a4 | Aged surface | 1 | 0 | 2 | Fractured |
| a5 | Aged surface | 10 | 2 | 15 | |
| b1 | Reverse surface | 0 | 2 | 2 | |
| b2 | Reverse surface | 0 | 5 | 3 | Fractured |
| b3 | Reverse surface | 0 | 7 | 6 | |
| b4 | Reverse surface | 0 | 3 | 3 | |
| b5 | Reverse surface | 0 | 3 | 3 | |

## E. Two-Layer Beam Model of the Neutral Surface and the Moment–Curvature Relation

To examine whether the through-thickness modulus contrast of the aged specimen can affect the bending response, we model the specimen as a rectangular beam composed of two perfectly bonded layers of equal thickness $h/2$ with different Young's moduli, representing the stiffer aged layer and the softer less-aged layer. We adopt the following assumptions: (i) plane cross-sections remain plane and normal to the neutral surface (Euler–Bernoulli bending), so the axial strain varies linearly through the thickness; (ii) the deflection is small and the curvature at mid-span is uniform, so that $R \gg h$; (iii) each layer is linearly elastic and isotropic, and the interface is perfectly bonded, so the strain is continuous across it; (iv) a uniaxial stress state is assumed (Poisson and width-direction coupling neglected); and (v) the modulus step is placed at the mid-plane as the simplest representation.

Here, $z$ is the through-thickness coordinate measured from the upper surface $(0 \leq z \leq h)$, and $h$ and $b$ are the thickness and width of the beam, respectively. $R$ is the radius of curvature of the neutral surface at mid-span. The local modulus is $E(z)$, with $E(z) = E_1$ for $0 \leq z < h/2$ and $E(z) = E_2$ for $h/2 < z \leq h$. Here $\bar{E} = (E_1 + E_2)/2$ is the mean modulus and $\Delta E = E_2 - E_1$ is the modulus difference between the layers (interchanging the layers, i.e., flipping the specimen, corresponds to $\Delta E \to -\Delta E$). The position of the neutral surface (zero-strain plane) measured from the upper surface is $\xi$, and $\varepsilon(z)$ and $\sigma(z)$ are the axial strain and stress, respectively. Finally, $dT$ is the axial force carried by the element of area $b\,dz$, and $M = Pl/4$ is the mid-span bending moment.

With the neutral surface at $z = \xi$, the linear strain field is

$$\varepsilon(z) = (\xi - z)/R,$$

the corresponding stress is $\sigma(z) = E(z)\,\varepsilon(z)$, and the axial force on the element $b\,dz$ is

$$dT = \sigma(z) b\,dz = (b/R) E(z)(\xi - z)\,dz.$$

In pure bending there is no net axial force, $T = 0$:

$$T = \int_0^h dT = \int_0^{h/2} \frac{bE_1}{R}(\xi - z)dz + \int_{h/2}^h \frac{bE_2}{R}(\xi - z)dz. \tag{A2}$$

Solving for $\xi$ gives $\xi = (h/4)(E_1 + 3E_2)/(E_1 + E_2)$, i.e., Eq. (5). The neutral surface is displaced from the mid-plane toward the stiffer layer by $(h/8)(\Delta E/\bar{E})$.

Taking moments about the neutral surface,

$$\begin{aligned} M &= -\int_0^h (\xi - z)dT = \int_0^h \frac{bE(z)}{R}(\xi - z)^2 dz \\ &= \int_0^{h/2} \frac{bE_1}{R}(\xi - z)^2 dz + \int_{h/2}^h \frac{bE_2}{R}(\xi - z)^2 dz. \end{aligned} \tag{A3}$$

Substituting Eq. (5) into Eq. (A3) and integrating it gives Eq. (6) in the main text.

This two-layer beam model has two important consequences. (i) Eq. (6) depends on the modulus difference only through $\Delta E^2$; therefore, interchanging the two layers (flipping the specimen, $\Delta E \to -\Delta E$) leaves the relation between $M$ and $R$ and hence between the load $P$ and the deflection $x$ unchanged. (ii) In the conversion of the raw data to bending stress-strain data, the neutral surface is assumed to lie at the mid-plane, so the outer-fiber distance $h/2$ is used for both the stress (Eq. (1)) and the strain (Eq. (2)). The apparent modulus is then $E_{\mathrm{app}} = \sigma/\varepsilon = MR/I = (12MR)/(bh^3) = \bar{E}[1 - (3/16)(\Delta E/\bar{E})^2]$, which is independent of orientation and equals the mean modulus $\bar{E}$ to second order in $\Delta E/\bar{E}$. A modulus contrast between the two layers thus cannot account for the difference between the two loading configurations.

## Author Information

### Corresponding Authors

Takato Ishida - Department of Materials Physics, Nagoya University, Furo-cho, Chikusa, Nagoya 464-8603, Japan; orcid.org/0000-0003-3919-2348; E-mail: ishida@mp.pse.nagoya-u.ac.jp

## Authors

Kazuya Haremaki - Department of Materials Physics, Nagoya University, Furo-cho, Chikusa, Nagoya 464-8603, Japan; orcid.org/0009-0007-8494-8008; E-mail: haremaki.kazuya.w2@s.mail.nagoya-u.ac.jp

Yusuke Koide - Graduate School of Engineering Science, The University of Osaka, 1-3 Machikaneyama, Toyonaka, Osaka 560-8531, Japan; orcid.org/0000-0002-4843-6888; E-mail: y.koide.es@osaka-u.ac.jp

Takashi Uneyama - Department of Materials Physics, Nagoya University, Furo-cho, Chikusa, Nagoya 464-8603, Japan; orcid.org/0000-0001-6607-537X; E-mail: uneyama@mp.pse.nagoya-u.ac.jp

Yuichi Masubuchi - Department of Materials Physics, Nagoya University, Furo-cho, Chikusa, Nagoya 464-8603, Japan; orcid.org/0000-0002-1306-3823; E-mail: mas@mp.pse.nagoya-u.ac.jp

## Conflict of Interest

The authors declare no competing interests.

## ACKNOWLEDGMENT

This work was supported by JSPS KAKENHI Grant Numbers 24K20949, CCI holdings Co., Ltd., Fuji Seal Foundation, CASIO Science Promotion Foundation, Fujikura foundation, and Tokai Pathways to Global Excellence (T-GEx), part of MEXT Strategic Professional Development Program for Young Researchers.